\documentclass[preprint,showpacs,preprintnumbers,amsmath,amssymb]{revtex4}

\usepackage{graphicx}
\usepackage{dcolumn}
\usepackage{bm}

\begin{document}

\preprint{APS/123-QED}

\title{Anisotropic dynamics in a shaken granular dimer gas experiment}

\author{J. Atwell}
\author{J. S. Olafsen}%
 \email{jolafsen@ku.edu}
\affiliation{Department of Physics and Astronomy, University of Kansas Lawrence, KS 66045}%

\date{\today}

\begin{abstract}
The dynamics, velocity fluctuations, and particle-plate interactions
for a 2D granular gas of shaken, non-spherical particles are studied 
experimentally.  The experiment consists of a horizontal plate that 
is vertically oscillated to drive the dynamics of macroscopic dimers,
spherical pairs that are loosely connected by a rod that couple the 
interaction each of the spheres has with the shaking plate.  The 
extended nature of the particles results in more than one energy-momentum 
transfer between the plate and each dimer per shaking cycle.  This 
complex interaction results in anisotropic behavior for the dimer that 
is a function of the shaking parameters. 
\end{abstract}

\maketitle


Recently, a novel driven granular gas experiment comprised of two layers of
separate species of granular material has demonstrated robust velocity 
fluctuations that are nearly Gaussian over a wide range of parameters
\cite{BaxterO}.  In the experiment, a horizontal plate vertically drives a
first layer of one species of particles whose collisions, in turn, fluidize a
second layer comprised of a different species of particles.  While the 
velocity fluctuations in the first layer, driven by the plate, remain strongly 
non-Gaussian and vary with the driving parameters, the second layer, 
thermalized by collisions with the first layer, demonstrate nearly Gaussian
velocity statistics over a wide range of shaking parameters.  Similar to this
mechanically fluidized bed, a gas fluidized bed system used to thermalize the
motion of a larger sphere also demonstrates nearly Maxwell-Boltzmann statistics
for the indirectly driven particle \cite{Durian}.  Clearly, the way in which
energy is injected into each layer in the mechanically fluidized experiment is
an important detail that must be examined more closely to understand how each
layer could simultaneously demonstrate drastically different velocity
fluctuations.

In this report, we detail the injection of energy between the shaking plate
and the species that comprise the lower layer of the two layer experiment: 
granular dimers, spherical pairs loosely connected by a short rod.
Experiments to probe the dynamics of driven granular gases, collections of 
large numbers of dissipative macroscopic particles whose motion is 
maintained through various forms of external driving, have focused largely 
on species of identical spheres 
\cite{Olafsen1,Wolfgang,Kudrolli,Aronson,Behringer,Warr}.  However, identical
spheres represent only one special species of macroscopic granular particles.
We demonstrate that 
many of the differences in the dynamic behavior from that of identical spheres
can be understood through anisotropies that directly relate to the geometry 
of the extended particles.  In 
particular, the manner in which the particles are thermalized by the shaking 
plate in the vertical direction is intimately connected to how the horizontal 
dynamics are manifested in the plane of the 2D granular gas.  A prior 
simulation has investigated developing a kinetic theory for such gases in
a freely cooling case \cite{Zippelius}.

The experiment consists of an aluminum plate of radius 14.6 cm that is 
vertically oscillated by an electromagnetic shaker.  The plate is level
and flat, and the acceleration of the plate, $\Gamma = A (2 \pi \nu)^2$, shaking
with peak amplitude, A, at frequency, $\nu$, is uniform, having a spatial 
variance across the surface that is less than 1$\%$.
The particles are 
comprised of two hollow spheres that are each 3.2 mm in diameter and are 
loosely connected by a thin rod that allows a spacing between the spheres of zero 
to 1.6 mm.  The total 
length of the dimer pair can therefore vary from 6.4 mm to 8.0 mm.  Except at 
highest densities where excluded volume effects compress the dimers along the 
interconnecting rod, the motion of the bouncing on the plate is observed to keep
the dimers fully extended.

The dimers
are constructed by cutting pairs from chains similar to those used in prior
investigations of the motion of chains \cite{BenNaim,polymerpaper,Wiggins}.  
The loose connection of the rod between the two spheres of the dimer allows for
an additional degree of movement where the spheres may rotate about the ends 
of the rod as wheels on an axle, letting the dimer roll in a direction 
perpendicular to the connecting rod.

The average mass of a single dimer is approximately 185 mg but can vary up to 
$\pm$ 7 mg from dimer to dimer.  The aspect ratio of the dimer, which is
the ratio of the length (as measured along the direction 
of the rod connecting the spheres) to the 
width, varies from 2 to 2.5 depending upon whether the dimer is fully 
compressed or fully extended.  A coefficient of restitution is determined
experimentally for the dimers to be approximately 0.3 \cite{coeff}.
For comparison, the identical technique yields a value of 0.5 for one of the 
single hollow balls from a dimer.  
A delrin ball is measured with the same technique to have a value of 0.9.

\begin{figure}[ht]
\includegraphics[height=50mm]{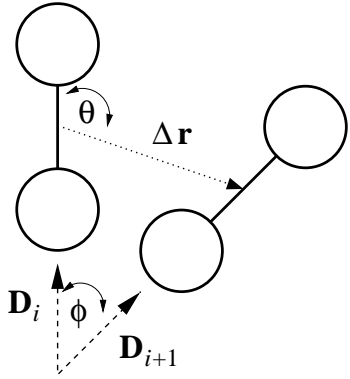}
\caption{\label{fig:one}When viewed from above, the horizontal dynamics of the non-spherical
 particles are described by two angles.  The angular displacement of the 
center of mass for the particle moving $\Delta {\bf r}$ relative to the
 natural director of
 the particle made by the connecting rod is $\theta$.   
The motion about the center of mass
 of the particle, $\phi$, denotes how the orientation of the director moves 
relative to a fixed frame.}
\end{figure}

The rod
connecting the two spheres is a natural geometric axis for the purpose of 
describing the motion of the particles.  When viewed from above,
the two dimensional 
motion of the dimer in the horizontal plane can be considered in 
terms of two angles, as shown in Figure \ref{fig:one}.  Because
of the extended
nature of the particle, the dynamics can be decomposed into the 
motion of the center of mass, $\theta$ and $\Delta {\bf r}$, and the  
motion about the center of mass, $\phi$.

The motion of a single dimer on a 
vertically shaken horizontal plate has been measured using high-speed digital
photography \cite{dalsacam}.  By overlaying two sequential images, a composite 
picture similar to Figure \ref{fig:one} can be produced.  The locations of the
two spheres that comprise the dimer can be identified using 
software analysis\cite{idl}.  Identifying the location of the spheres that 
constitute the dimer in each frame allows the center of mass and the orientation 
of the dimer to be determined.  If the orientation of the dimer relative to a
fixed frame (such as that of the camera) is {\bf D}$_i$ and {\bf D}$_{i+1}$ 
in frame i and i+1, respectively, then the angular displacement about the 
center of mass $\phi$ is determined from the following relationship:
\begin{equation}
sin(\phi) = \frac{{\bf D}_i \times {\bf D}_{i+1}}{\mid{\bf D}_i\mid \mid{\bf D}_{i+1}\mid}.
\label{eq:one}
\end{equation}
\noindent
In a similar fashion, the direction of the motion of the center of mass can be
determined by examining the motion of the center of mass, $\Delta${\bf r}, from
one frame to the next relative to the director of the dimer, {\bf D}$_i$:
\begin{equation}
cos(\theta) = \frac{{\bf D}_i \cdot \Delta {\bf r}}{\mid{\bf D}_i\mid \mid \Delta {\bf r} \mid}.
\label{eq:two}
\end{equation}

\begin{figure}[ht]
\includegraphics[height=50mm]{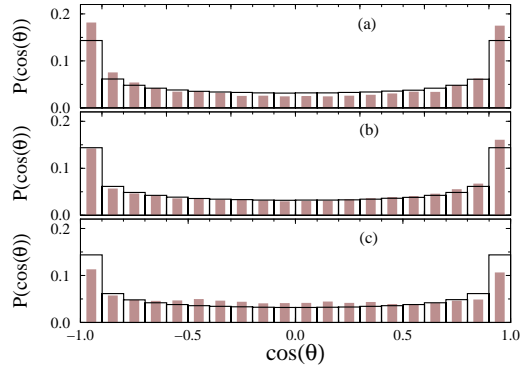}
\caption{\label{fig:two}Histograms (dark boxes) for $cos(\theta)$, the direction of center of mass motion of a 
single dimer on a plate with a peak acceleration of 2g for (a) 50 Hz, (b) 70 Hz, 
and (c) 90 Hz.  The distribution for $cos(\theta)$ for a uniform distribution of $\theta$ is shown by the open boxes in each figure.  The distribution demonstrates a bias for the dimer to move in a 
direction along the director rod joining the two spheres of the dimer at low frequency.  The bias is
shifted to motion perpendicular to the connecting rod as the shaking amplitude is decreased (shaking frequency is increased).}
\end{figure}

Probability distributions of $cos(\theta)$ for a single dimer on a plate for 
different shaking frequencies at a peak plate acceleration of 2g are shown in 
Figure \ref{fig:two}.  The particle was 
tracked from above with a high speed camera and the results demonstrate a preference for 
the center of mass of the dimer to move along the direction of the rod connecting 
the two spheres at low frequencies.  The reader should note the change in this bias by the decrease
in the probability at $cos(\theta) = \pm 1$ as the shaking frequency is increased.
There is an interesting exchange between the motion of the center of mass and about the center of
mass as the frequency is increased.  At 50 Hz, when the motion about the center of mass is
biased toward the ends, the center of mass motion demonstrates Gaussian velocity statistics.  As 
the frequency of shaking is increased to 70 Hz and the motion about the center of mass becomes more uniform,
the velocity statistics for the center of mass become slightly non-Gaussian.  At 90 Hz, where the motion about
the center of mass is now biased for motion perpendicular to the director, the velocity statistics for the 
center of mass motion is strongly non-Gaussian. 
To understand this anisotropy in the dimer motion, it is necessary to examine how the single dimer interacts with the 
vertically shaking plate.

\begin{figure}[ht]
\includegraphics[height=50mm]{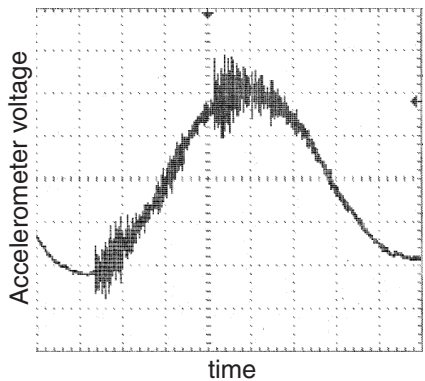}
\caption{\label{fig:three}Accelerometer signal for a single dimer chattering on the plate.  The vertical scale is 100 millivolts (1g) per major division.  The horizontal scale is 2.5 milliseconds per major division.  The driving frequency was 50 Hz and the acceleration amplitude was 2g.}
\end{figure}

\begin{figure}[ht]
\includegraphics[height=50mm]{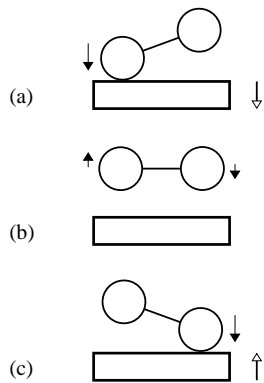}
\caption{\label{fig:four}Viewed from the side, the dimer's motion is biased by the relative motion
of the shaking plate during each collision.  In (a), the one sphere of the dimer collides while the plate is 
accelerating downward. In this collision, the plate absorbs some of the momentum of the ball that recoils with a 
slower velocity as the second sphere begins to fall toward the plate (b).  
In (c), the second sphere of the dimer collides while the plate is accelerating upward, this collision imparting a momentum gain to the dimer.}
\end{figure}

Figure \ref{fig:three} is an example of the signal acquired from the 
accelerometer to measure the peak acceleration of the shaking plate in this experiment.  
The data shown is for one typical oscillation of the plate with a peak acceleration of 2g at a 
frequency of 50 Hz.  The signal is comprised of
two parts.  The sinusoidal wave is the measured acceleration of the plate.  The 
two ``noisy'' portions of the signal (one near the trough and one near the peak of
the oscillation) are the pings caused by each of the balls of the dimer colliding 
with the plate.  The trough (negative acceleration) ping corresponds to a ball of
the dimer colliding with the plate as it accelerates downward (see also Figure \ref{fig:four}(a)).  
The peak ping corresponds to the other ball of the dimer colliding with the plate 
as it accelerates upward (Figure \ref{fig:four}(c)).    
The bias in Figure \ref{fig:two} can be understood in light of this dimer-plate
interaction in the vertical direction in terms of the individual collisions of each 
sphere of the dimer.

The motion of a single sphere interacting with a shaking plate has been studied previously \cite{luck}.
In carefully examining the motion of the single sphere throughout the shaking cycle, regions of the system's
phase space were observed to either increase or decrease the sphere's momentum.  These regions were deemed
transmitting and absorbing, respectively \cite{luck}, and were responsible for the general ``chattering''
up and down in the amplitude of the sphere's motion over several shaking cycles.  The analysis for the motion of the
dimer may be interpreted for two spheres colliding at nearly opposite phases of the plate cycle.

When the one ball of the dimer collides with the 
plate during its downward acceleration, the dimer loses momentum (as would a single sphere in that phase of the 
shaking cycle \cite{luck}) in the direction along the rod (Figure \ref{fig:four}(a)).
The first ball of the dimer recoils from this ``soft'' collision with the plate with a reduced velocity as the 
other ball of the dimer, under the influence of gravity, begins to fall toward the plate (Figure \ref{fig:four}(b)).  
When the plate is accelerating upward, the dimer gains a momentum kick from the second sphere
colliding with the plate (Figure \ref{fig:four}(b)), as would a single sphere under the same circumstances
\cite{luck}.  That momentum kick is transmitted up the loose connection of the rod of the dimer.  
The result is a net momentum transfer along the rod of the dimer.  
This net interaction with the shaking plate results
in a bias of the particle to move in a direction along the geometric axis of the 
dimer.  The biasing effect grows at fixed acceleration as the frequency is decreased because 
the amplitude of shaking increases, increasing the vertical angle the dimer makes relative
to the horizontal plane and emphasizing the bias between the two collisions.  As the
amplitude of shaking decreases, this angular rotation decreases and the effect is
minimized but never completely disappears (see Figure \ref{fig:two}).  
It should also be noted that this net momentum transfer led to
a second important difference from the dynamics of monomers 
\cite{Olafsen1,Olafsen2}.  In
the monomer case, clustering could be avoided within the system by shaking at an 
acceleration as little as 1.25g for some values of A and $\nu$.  Because of the more complicated
momentum transfer between the dimers and the plate, accelerations of $\approx 4g$ were
necessary to produce a nearly uniform density gas of dimers, i.e. to avoid clustering.

\begin{figure}[ht]
\includegraphics[height=50mm]{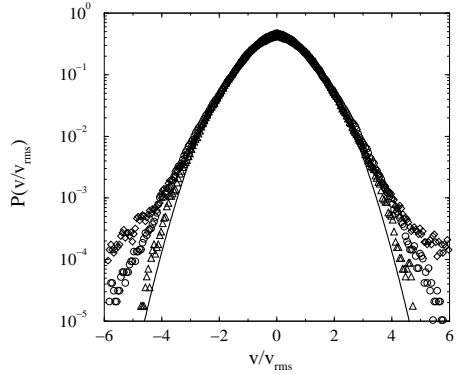}
\caption{\label{fig:five}Velocity fluctuations for different shaking parameters.  
Triangles:  Frequency is 40 Hz, $\Gamma$ = 1.65g, density = 0.27, $v_{rms}$ = 1.74 cm/s.  
Circles:  Frequency is 65 Hz, $\Gamma$ = 2.0g , density = 0.87, $v_{rms}$ = 1.04 cm/s.
Diamonds:  Frequency is 65 Hz, $\Gamma$ = 1.5g, density = 0.55, $v_{rms}$ = 0.94 cm/s.}
\end{figure}

For more than one dimer on the plate, the accelerometer signal becomes more 
complicated and it is difficult to determine which particle is interacting with the
plate.  However, in a future paper we will demonstrate a method of describing the
net momentum transfer for several dimers interacting with the plate to show that collisions occur
more uniformly throughout the shaking cycle than in the monomer case.  Our goal in the current work 
is to describe the velocity statistics of an inelastic granular gas comprised 
of many dimers (800 - 3000) as measured via high speed photography from above the plate for 
motion in the horizontal plane.      
The results for some parameters are shown in Figure \ref{fig:five}.  
In nearly all regimes, the velocity distributions demonstrate non-Gaussian behavior 
similar to that of the monomer experiment of Olafsen and Urbach \cite{Olafsen1,Olafsen2}.  
It is interesting to note, however, that at the lower density, the velocity distribution 
function is nearly Gaussian.  This can also be understood in light of the complex interaction
the particles have with the plate.

The density of the dimer gas, or coverage of the plate, is calculated in terms of 
the equivalent number of dimer balls that would be needed to have a single layer 
on the shaking plate.  This number is 7320 spheres or 3660 dimers 
(assuming full compression of all dimers along their 
connecting rods for one full layer).  
The density is then calculated 
as a fraction of the number of dimers in the cell normalized by 3660. 

At low densities, the dimers do not undergo as 
many collisons with the other particles in the gas and the horizontal motion is
dominated by local surface roughness of the plate, resulting in the nearly Gaussian
distributions in the horizontal direction. However, the dimers still interact, so the motion
cannot completely be thought of as that of a single isolated dimer as was the case in 
Figure ~\ref{fig:two}.
Indeed, at higher densities, the horizontal motion is dominated by dimer-dimer collisions, 
which in general can be thought of belonging to one of three types.  
Because each dimer has two interactions with the shaking plate as previously described, each
dimer can be thought of as a pair of loosely coupled spheres, one for which there was a net momentum
gain, $p_+$, and one for which there was a net momentum loss, $p_-$, in its most recent interaction with
the shaking plate.

Collisions between any two dimers $i$ and $j$ are therefore interactions of their constituent spheres 
that have most recently both received positive momentum kicks, $<p^{i}_+ p^{j}_+>$, 
both received negative momentum kicks, $<p^{i}_- p^{j}_->$, or two spheres that have received one of each type 
of kick from interacting with the plate, $<p^{i}_+ p^{j}_->$ or $<p^{i}_- p^{j}_+>$.  
The way in which energy and 
momentum are transferred from the vertical to the horizontal is therefore much more
complicated to model.  In general, as the density of the dimer gas increases, the velocity
statistics become more non-Gaussian.  However, because of the additional complexity of dimer collisions, it is not 
surprising that the overall shape of the deviations from Guassian statistics do not
mimic the monomer case, where the shape of the entire distribution  
could be ``tuned'' from nearly Gaussian to nearly exponential depending on the shaking
parameters and how ``two dimensional'' the system was constrained to be \cite{Olafsen2}.  The 
other general trend observed in the data is that the deviation from Gaussian behavior 
increases as the root mean square velocity of the inelastic gas decreases.

In conclusion, in this paper we detail similarities and differences in  
the observed dynamic behavior of a driven granular gas composed of dimers (linked 
spheres) from that of granular gases with monomer species.  The dynamic differences
are directly related to the geometric anisotropies of the constituent particles and
the additional complexity with which these particles interact with the shaking plate.
The results clearly underscore the need to better understand how energy and momentum
are injected in such systems in order to study the dynamics of driven granular gases. 


\begin{acknowledgments}
This work was supported by a grant from the Petroleum Research Fund of the 
American Chemical Society and a grant from the General Research Fund of the
University of Kansas.  One of us (JA) was also partially supported by an
Undergraduate Research Award (UGRA) from KU.  The authors
would like to thank G. William Baxter for several helpful conversations and feedback
in the preparation of this manuscript. 
\end{acknowledgments}


\bibliography{dimerpap}

\begin{thebibliography}{17}
\bibitem[{dal()}]{dalsacam}
\bibinfo{note}{A 256x256 pixel, 955 frame per second (fps) Dalsa-CA-D6 8-bit
  camera was used.}
\end{thebibliography}





\end{document}